# Structured low-rank matrix completion for forecasting in time series analysis

Jonathan Gillard[a], Konstantin Usevich[b,c,*]

[a]*Cardiff School of Mathematics, Cardiff University, Senghennydd Road, Cardiff, UK, CF24 4AG.*
[b]*Université de Lorraine, CRAN, UMR 7039, Campus Sciences, BP 70239, 54506 Vandœuvre-lès-Nancy, France*
[c]*CNRS, CRAN, UMR 7039, France*

**Abstract**

In this paper we consider the low-rank matrix completion problem with specific application to forecasting in time series analysis. Briefly, the low-rank matrix completion problem is the problem of imputing missing values of a matrix under a rank constraint. We consider a matrix completion problem for Hankel matrices and a convex relaxation based on the nuclear norm. Based on new theoretical results and a number of numerical and real examples, we investigate the cases when the proposed approach can work. Our results highlight the importance of choosing a proper weighting scheme for the known observations.

## 1. Introduction

Low-rank representations and approximations have been shown to be a very useful tool in time series forecasting [1]. One of the popular approaches is singular spectrum analysis (SSA) forecasting [2] that embeds the time series into a Hankel matrix and uses a low-rank approximation and continuation to compute the next values of a time series. SSA uses the fact that many time series can be well approximated by a class of so-called time series of finite rank. Despite many successful examples [3, 4, 5], SSA forecasting has a number of disadvantages.

In this paper we develop a method based on Hankel matrix completion. We follow the approach of [6], where it was proposed to embed a time series into a Hankel matrix and the missing data (to be forecasted) are stored in the bottom right-hand corner of this matrix. The method of [6] is based on minimising the nuclear norm, which provides a convex relaxation to a low-rank matrix completion problem which is non-convex and NP-hard in general (see for example [7] and [8]).

The nuclear norm (the sum of singular values) is a popular convex surrogate for the rank [7], and is similar to using the $\ell_1$-norm for sparse approximation [9]. It was shown to be a successful tool in imputing the missing values of a matrix (see for example [7], [10], [11] and [12], [13]). Nuclear norm relaxation has been a very popular tool for spectral estimation [13], recommender systems [7], and system identification [14, 15, 16]. An advantage of the nuclear norm relaxation

*Corresponding author
Email addresses:* GillardJW@Cardiff.ac.uk (Jonathan Gillard), konstantin.usevich@univ-lorraine.fr (Konstantin Usevich)

considered in this paper is the ability to build more complex models to represent potentially more complex behavior in observed time series.

An important question is when the convex relaxation solves the original low-rank matrix completion problem. Much famous research has been conducted on this topic, but most available research [7, 10, 11, 12, 13] assumes that the position of the missing entries in the matrix is random, and often that the known entries are also random; mostly unstructured matrices are considered. Therefore these results are not applicable due to a special arrangement of missing data and due to the Hankel structure in our problem. Moreover, as noted in [17] and [18], the case of structured matrices is much more challenging.

There are few available results for completion of Hankel matrices with a fixed pattern of missing values. In [19], a special case was analysed: square real-valued Hankel matrices with nearly half of its values missing. It was shown that the nuclear norm relaxation gives the correct rank-one completion only when the embedded time series can be written as a sum of decreasing exponentials. In [20], this analysis was extended to the rank-$r$ case for the same pattern of missing values.

This paper has several contributions. First, as in [6], we consider the general case of rectangular Hankel matrices with potentially fewer missing values. We show that for few missing values, the convex relaxation of the low-rank matrix completion using the nuclear norm will give identical solutions without using the convex relaxation for time series with undamped or exponentially increasing periodic components, and establish bounds on the number of missing values. We also study the question of choosing the optimal shape of the Hankel matrix (parameterized by the so-called window length). Second, we suggest a new (relative to [6]) formulation of the low-rank matrix completion problem for Hankel matrices, which allows the possibility to allocate past observations different weights. In particular, exponential weighting is designed to overcome the problems related to the performance of the nuclear norm for time series that can be expressed as a sum of increasing exponentials.

Empirical comparisons show that, with the proper choice of weights, our novel formulation performs well relative to a number of classical techniques. For the numerical examples in this paper we use CVX, a MATLAB package for specifying and solving convex programs [21, 22].

This paper has the following structure. In Section 2 we formally define the problems to be considered. We first define exact matrix completion, before considering an approximate version. This section also describes the settings used throughout the paper. Some known theoretical results necessary to be stated are reviewed in Section 3. First, the time series of finite rank are recalled, and the solution of the exact minimal rank completion is summarized. Next, known results on time series of finite rank are recalled. Section 4 contains the main results of the paper. First, we give theoretical bounds for matrix completion in the case of arbitrary shape of the matrix and number of missing values. We check the tightness of our bounds based on numerical experiments. Second, we establish the connection between exponential weighting and preprocessing of time series. Finally, the examples for forecasting of real and model time series demonstrating the advantages of the proposed methodology are provided in Section 5.

## 2. Problem Statement

### 2.1. Hankel matrices

For a vector $\mathbf{f} = (f_1, \ldots, f_n)$ with $n > 1$ and a so-called window length $L$, the $L \times (n - L + 1)$ Hankel matrix is defined as

$$\mathscr{H}_L(\mathbf{f}) = \begin{pmatrix} f_1 & f_2 & \cdots & f_{n-L+1} \\ f_2 & f_3 & \cdot^{\cdot^{\cdot}} & f_{n-L+2} \\ \vdots & \cdot^{\cdot^{\cdot}} & \cdot^{\cdot^{\cdot}} & \vdots \\ f_L & f_{L+1} & \cdots & f_n \end{pmatrix}.$$

In what follows, we are going to pose the problem of forecasting a given time series as the low-rank matrix completion of a Hankel matrix. Formally, let

$$\mathbf{p} = (p_1, p_2, \ldots, p_{n+m}), \tag{1}$$

be a vector of length $(n + m)$, with $m \geqslant 0$. In what follows $m$ will be the number of observations forecasted, and $n$ will be the length of the given time series that we wish to forecast. We use the notation $\mathbf{p}_{(1:n)} = (p_1, p_2, \ldots, p_n)$ for the first $n$ elements of $\mathbf{p}$. Next, let $L$, $K$ be the integers such that $L + K - 1 = m + n$. Then the matrix structure $\mathcal{S}(\mathbf{p})$ (parameterized by $\mathbf{p}$) we consider is

$$\mathcal{S}(\mathbf{p}) = \mathscr{H}_L(\mathbf{p}) = \begin{pmatrix} p_1 & p_2 & \cdots & \cdots & \cdots & p_K \\ p_2 & p_3 & \cdots & \cdot^{\cdot^{\cdot}} & \cdots & \vdots \\ \vdots & \cdots & \cdot^{\cdot^{\cdot}} & \cdot^{\cdot^{\cdot}} & \cdots & p_n \\ \vdots & \cdot^{\cdot^{\cdot}} & \cdot^{\cdot^{\cdot}} & \cdot^{\cdot^{\cdot}} & \cdots & p_{n+1} \\ \vdots & \cdot^{\cdot^{\cdot}} & \cdot^{\cdot^{\cdot}} & \cdot^{\cdot^{\cdot}} & \cdot^{\cdot^{\cdot}} & \vdots \\ p_L & \cdots & p_n & p_{n+1} & \cdots & p_{n+m} \end{pmatrix}. \tag{2}$$

In (2), the grey-shaded values are "known" and others are "missing".

The Hankel matrix structure belongs to the class of affine matrix structures [18, §3.3] having the form:

$$\mathcal{S}(\mathbf{p}) = \mathbf{S}_0 + \sum_{i=1}^{(n+m)} p_i \mathbf{S}_i \tag{3}$$

where $\mathbf{S}_i$, $i \in \{0, 1, \ldots, (n+m)\}$ are given linearly independent basis matrices, and in particular, for the Hankel matrix structure (2), the basis matrices matrices in (3) are given as $\mathbf{S}_0 = 0$,

$$\mathbf{S}_1 = \begin{pmatrix} 1 & 0 & \cdots & 0 & 0 \\ 0 & 0 & \cdot^{\cdot^{\cdot}} & 0 & 0 \\ \vdots & \cdot^{\cdot^{\cdot}} & \cdot^{\cdot^{\cdot}} & \cdot^{\cdot^{\cdot}} & \vdots \\ 0 & 0 & \cdot^{\cdot^{\cdot}} & 0 & 0 \\ 0 & 0 & \cdots & 0 & 0 \end{pmatrix}, \mathbf{S}_2 = \begin{pmatrix} 0 & 1 & \cdots & 0 & 0 \\ 1 & 0 & \cdot^{\cdot^{\cdot}} & 0 & 0 \\ \vdots & \cdot^{\cdot^{\cdot}} & \cdot^{\cdot^{\cdot}} & \cdot^{\cdot^{\cdot}} & \vdots \\ 0 & 0 & \cdot^{\cdot^{\cdot}} & 0 & 0 \\ 0 & 0 & \cdots & 0 & 0 \end{pmatrix}, \ldots, \mathbf{S}_{n+m-1} = \begin{pmatrix} 0 & 0 & \cdots & 0 & 0 \\ 0 & 0 & \cdot^{\cdot^{\cdot}} & 0 & 0 \\ \vdots & \cdot^{\cdot^{\cdot}} & \cdot^{\cdot^{\cdot}} & \cdot^{\cdot^{\cdot}} & \vdots \\ 0 & 0 & \cdot^{\cdot^{\cdot}} & 0 & 1 \\ 0 & 0 & \cdots & 1 & 0 \end{pmatrix}, \mathbf{S}_{n+m} = \begin{pmatrix} 0 & 0 & \cdots & 0 & 0 \\ 0 & 0 & \cdot^{\cdot^{\cdot}} & 0 & 0 \\ \vdots & \cdot^{\cdot^{\cdot}} & \cdot^{\cdot^{\cdot}} & \cdot^{\cdot^{\cdot}} & \vdots \\ 0 & 0 & \cdot^{\cdot^{\cdot}} & 0 & 0 \\ 0 & 0 & \cdots & 0 & 1 \end{pmatrix}.$$

In the following subsections, we describe the formal statement of the matrix completion problem, and the nuclear norm relaxation approach that we propose.

## 2.2. Exact low-rank matrix completion

Let $\mathbf{p}_0 = (p_{0,1}, p_{0,2}, \ldots, p_{0,n})$ be a given vector of observations (a time series). For a given matrix structure (3), the Structured Low-Rank Matrix Completion (SLRMC) problem is posed as

$$\tilde{\mathbf{p}} = \underset{\mathbf{p} \in \mathbb{R}^{(n+m)}}{\arg\min} \operatorname{rank} \mathcal{S}(\mathbf{p}) \text{ subject to } \mathbf{p}_{(1:n)} = \mathbf{p}_0. \tag{4}$$

The implicit low-rank assumption of the Hankel matrix corresponds to the class of time series of so-called finite rank, which are described in Section 3.

The matrix completion problem (4) for general matrix structures is NP-hard (see [7] and [8]). A convex relaxation of this problem based on the nuclear norm became increasingly popular recently. Formally, for a matrix $\mathbf{X} \in \mathbb{C}^{L \times K}$ its nuclear norm is defined as

$$\|\mathbf{X}\|_* = \sum_{k=1}^{\min(L,K)} |\sigma_k(\mathbf{X})|,$$

where $\sigma_k(\mathbf{X})$ are the singular values of $\mathbf{X}$. A convex relaxation of (4) is obtained by replacing the rank with the nuclear norm:

$$\hat{\mathbf{p}} = \underset{\mathbf{p} \in \mathbb{R}^{(n+m)}}{\arg\min} \|\mathcal{S}(\mathbf{p})\|_* \text{ subject to } \mathbf{p}_{(1:n)} = \mathbf{p}_0. \tag{5}$$

The intuition behind this relaxation is the same as for using the $\ell_1$ norm in compressed sensing: the nuclear norm is expected to force all but a few singular values to be zero (a low-rank solution).

**Remark 2.1.** *For the Hankel matrix case, the solution of (4) is known and is given in Section 3. Still, the performance of the nuclear norm relaxation is important for understanding the behavior of forecasting in the approximate case to be introduced in the next subsection.*

## 2.3. Approximate matrix completion

Let $\mathbf{x} = (x_1, x_2, \ldots, x_n)$ be a vector of length $n$. We denote

$$\|\mathbf{x}\|_W = \sqrt{\sum_{i=1}^{n} w_i x_i^2} \tag{6}$$

where $W = (w_1, w_2, \ldots, w_n)$, is a vector of weights, $w_i > 0$ for $i = 1, \ldots, n$.

The approximate rank minimization can be posed as follows. Given $\mathbf{p}_0$, $W$, $m \geqslant 0$ and $\tau \geqslant 0$ find

$$\tilde{\mathbf{p}} = \underset{\mathbf{p} \in \mathbb{R}^{(n+m)}}{\arg\min} \operatorname{rank} \mathcal{S}(\mathbf{p}) \text{ subject to } \|\mathbf{p}_{(1:n)} - \mathbf{p}_0\|_W \leqslant \tau. \tag{7}$$

The parameter $\tau$ controls the precision of approximation. Two extreme cases can be distinguished:

- if $\tau = 0$, then (7) is equivalent to the exact matrix completion problem (4), i.e. there is no approximation;

- if $m = 0$ in (7), then $\tilde{\mathbf{p}}$ is an approximation to the given vector $\mathbf{p}_0$ with no forecast.

Unlike (4), the problem (7) does not have a known solution. In fact, it is a dual problem to structured low-rank approximation [18] which is known to be a difficult optimization problem [23].

In order to circumvent the complexity of the problem, we consider the following relaxation of (7), using the nuclear norm:

$$\mathbf{p}_* = \underset{\mathbf{p}\in\mathbb{R}^{(n+m)}}{\arg\min} \|\mathcal{S}(\mathbf{p})\|_* \text{ subject to } \|\mathbf{p}_{(1:n)} - \mathbf{p}_0\|_W \leq \tau. \tag{8}$$

**Remark 2.2.** *There are alternative ways to extend the problem (5) to approximate versions. For example, consider the following equivalent formulations:*

$$\min_{\mathbf{p}\in\mathbb{R}^{(n+m)}} \|\mathbf{p}_{(1:n)} - \mathbf{p}_0\|_W \text{ subject to } \|\mathcal{S}(\mathbf{p})\|_* \leq \delta, \tag{9}$$

$$\min_{\mathbf{p}\in\mathbb{R}^{(n+m)}} \|\mathbf{p}_{(1:n)} - \mathbf{p}_0\|_W + \gamma\|\mathcal{S}(\mathbf{p})\|_*, \tag{10}$$

*where $\delta$ and $\gamma$ are regularisation parameters for each of the formulations.*

*In fact, it can be shown that (similarly to the result for equivalence of LASSO formulations [24]) the problems (8), (9) and (10) are equivalent in the following sense: for any value of $\tau$, there exist $\delta$ and $\gamma$ such that the solutions to (8), (9) and (10) coincide. However, the relation between "equivalent" $\tau$, $\delta$ and $\gamma$ is not known a priori. See for example [24, §1.3].*

### 2.4. Choice of weights

There are several natural choices of weights $W$ defining (6) in the approximation problem (8):

1. *Trapezoid weighting.* Take $W_1$ to be the vector of weights such that $\|\mathbf{p}_{(1:n)} - \mathbf{p}_0\|_{W_1} = \|\mathcal{H}_L(\mathbf{p}_{(1:n)}) - \mathcal{H}_L(\mathbf{p}_0)\|_F$. In this case, the vector $W_1 = (w_{1,1}, w_{1,2}, \ldots, w_{1,n})$ is given by

$$w_{1,i} = \begin{cases} i, & \text{for } i = 1, \ldots, L-1, \\ L, & \text{for } i = L, \ldots, n-L+1, \\ n-i+1, & \text{for } i = n-L+2, \ldots, n. \end{cases} \tag{11}$$

2. *Uniform weighting.* Take $W_2$ to be the vector of weights such that $W_2 = (1, 1, \ldots, 1)$.
3. *Exponential weighting.* take $W_3$ to be the vector of weights $W_3 = (w_{3,1}, w_{3,2}, \ldots, w_{3,n})$ such that $w_{3,i} = \exp(\alpha i), i = 1, \ldots, n$ for some number $\alpha$.

We make the following comments regarding each of the possible choices of weights given above. $W_1$ is the vector of weights 'enforced' upon the observations if the Frobenius norm of a Hankel matrix is used. The Frobenius norm for matrix optimization problems is commonly and traditionally used due to classical results of the optimality of low rank approximations achieved by truncating the singular value decomposition. However an unintended consequence of using the Frobenius norm is that the vector of observations (when placed into a Hankel matrix) receive the weights as given in $W_1$. Such weights may be unnatural for forecasting, for example, as the more recent observations are given declining weight in time. $W_2$ is the natural 'correction' to the weights offered in $W_1$ by giving each observation equal weight. As mentioned earlier, for the purposes of forecasting there may be some advantages in adopting the weights given in $W_3$, where the most recent observations receive larger weight.

## 3. Time series of finite rank

In this section, we recall the class of so-called time series of finite rank [2] and the solution of the exact rank minimization problem (4) for the Hankel matrix structure (2).

*3.1. Time series of finite rank and linear recurrent formulae*

First, we recall basic properties of time series of finite rank[1] [2]. Informally speaking, time series of finite rank are the time series for which the Hankel matrix has low rank. It is known [2, Chapter 5] that the class of such real-valued time series is given by sums of products of cosines, exponential and polynomial functions, i.e.

$$p_k = \sum_{j=1}^{s_1} Q_j(k)\rho_j^k \cos(2\pi\omega_j k + \phi_j) + \sum_{j=s_1+1}^{s_2} Q_j(k)\rho_j^k, \quad k = 1, 2, \ldots, \tag{12}$$

where $Q_j(k)$ are real polynomials of degrees $\mu_j - 1$ ($\mu_j$ are positive integers), $\omega_j \in (0, \frac{1}{2})$, $\phi_j \in \mathbb{R}$, $\rho_j \in \mathbb{R}$ are such that $\{\rho_j\}_{j=s_1+1}^{s_2}$ are distinct and the pairs $\{(\rho_j, \omega_j)\}_{j=1}^{s_1}$ are also distinct. As shown in [2], time series of the form (12) are particularly suitable to model trends, periodicities and modulated periodicities in time series analysis. The number

$$r = \sum_{j=1}^{s_1} 2\mu_j + \sum_{j=s_1+1}^{s_2} \mu_j \tag{13}$$

is called the finite difference dimension (or rank) of the time series and is equal to the rank of the Hankel matrix $\mathcal{S}(\mathbf{p})$ given in (2) if $L, K \geq r$ (see [2, Chapter 5]).

Time series of finite rank can be more compactly represented in the complex-valued case. Some of the results of this paper will be formulated for complex-valued time series. Next, we recall a summary from [25], which is based on the results of [26]. Consider a complex-valued infinite time series $(p_1, p_2, \ldots, p_n, \ldots)$ with

$$p_k = \sum_{j=1}^{s} P_j(k)\lambda_j^k, \quad k = 1, 2, \ldots, \tag{14}$$

where $P_j(k)$ are complex polynomials of degrees $\nu_j - 1$, ($\nu_j$ are positive integers), and $\lambda_j \in \mathbb{C}\setminus\{0\}$.

Note that time series (12) are special cases of time series (14). Indeed, take $s = s_1 + 2s_2$,

$$(\lambda_1, \ldots, \lambda_s) = (\rho_1 e^{j2\pi\omega_1}, \rho_1 e^{-j2\pi\omega_1}, \ldots, \rho_{s_1} e^{j2\pi\omega_{s_1}}, \rho_{s_1} e^{-j2\pi\omega_{s_1}}, \rho_{s_1+1}, \ldots, \rho_{s_1+s_2})$$

and $(P_1, \ldots, P_s) = (\frac{e^{j\phi_1}}{2}Q_1, \frac{e^{-j\phi_1}}{2}Q_1, \ldots, \frac{e^{j\phi_{s_1}}}{2}Q_{s_1}, \frac{e^{-j\phi_{s_1}}}{2}Q_{s_1}, Q_{s_1+1}, \ldots, Q_{s_1+s_2})$.

It is known [25, Corollary 2.1] that time series (14) satisfy the minimal linear recurrent formula

$$p_{k+r} = -\sum_{j=0}^{r-1} q_j p_{k+j}, \tag{15}$$

where the (complex) coefficients $q_j$ are the coefficients of the so-called characteristic polynomial

$$q(z) = (z - \lambda_1)^{\nu_1} \cdots (z - \lambda_s)^{\nu_s} = z^r + q_{r-1}z^{r-1} + \cdots + q_1 z + q_0. \tag{16}$$

This is why the number $r = \sum_{j=1}^{s} \nu_j$ is called the finite difference dimension of the time series (14).

---

[1] In fact, we consider the class of time series of finite difference dimension; however, it coincides with the time series of finite rank if the time series is infinite [2, Corollary 5.1]. This is why we use the term "time series of finite rank" in this paper.

*3.2. Solution of the rank minimization problem*

For time series of finite rank, the solution of the rank minimization problem (4) for the Hankel structure (2) is known. It is equivalent to the problem of minimal rank extension of Hankel matrices, solved in [27] for square matrices and in [26] for rectangular matrices. We summarize the results of [27, 26] in the form of a theorem, and the proof is given in Appendix B.

**Theorem 3.1.** *Let $r \leqslant \min(L, K, \frac{n}{2})$, and consider the Hankel matrix structure as given in (2).*

*Then for any complex-valued time series $(p_{0,1}, p_{0,2}, \ldots, p_{0,n}, \ldots)$ of finite rank $r$ (i.e. of the form (14)), and vector $\mathbf{p}_0$ defined as*

$$\mathbf{p}_0 = (p_{0,1}, p_{0,2}, \ldots, p_{0,n}), \qquad (17)$$

*the solution of the rank minimization problem*

$$\widetilde{\mathbf{p}} = \underset{\mathbf{p} \in \mathbb{C}^{(n+m)}}{\arg \min} \operatorname{rank} \mathcal{S}(\mathbf{p}) \text{ subject to } \mathbf{p}_{(1:n)} = \mathbf{p}_0 \,. \qquad (18)$$

*is unique and given by*

$$\widetilde{\mathbf{p}} = (p_{0,1}, p_{0,2}, \ldots, p_{0,n+m}) \,, \qquad (19)$$

*where $p_{0,n+1}, p_{0,n+2} \ldots p_{0,n+m}$ are computed using the linear recurrent formula (15).*

**Remark 3.2.** *Theorem 3.1 implies that the same holds for the real-valued case: for a real-valued time series $(p_{0,1}, p_{0,2}, \ldots, p_{0,n}, \ldots)$ of finite rank $r$ (i.e., time series of the form (12)) and the vector $\mathbf{p}_0$ defined as (17), the solution of the rank minimization problem (4) is unique and given by (19).*

Variants of Theorem 3.1 in the real-valued case are also used in the framework of SSA [2, 25]. In fact, the continuation by the linear recurrent formula (15) is at the core of forecasting methods in SSA.

**Remark 3.3.** *If the parametric form of the time series (i.e. (12) or (14)) is known, then the minimal rank completion is given by the same formula (15). However, the advantage of the recursion (15) is that the parametric form does not have to be derived. Due to that reason such matrix completion approaches are referred to as data-driven in [28]. The nuclear norm forecasting method proposed in this paper has the same advantage.*

*3.3. Performance of the nuclear norm: known results*

The performance of the nuclear norm (i.e. when the solution of (5) coincides with the solution of (4)), was studied recently for a special case of the structure (2):

$$\mathcal{S}(\mathbf{p}) = \begin{pmatrix} p_1 & p_2 & \cdots & p_n \\ p_2 & p_3 & \cdot^{\cdot^{\cdot}} & p_{n+1} \\ \vdots & \cdot^{\cdot^{\cdot}} & \cdot^{\cdot^{\cdot}} & \vdots \\ p_n & p_{n+1} & \cdots & p_{n+m} \end{pmatrix}, \qquad (20)$$

i.e. when the matrix is square ($L = K$), and all the values below the main antidiagonal of the Hankel matrix $\mathcal{S}(\mathbf{p})$ are missing (i.e. $L = n = m + 1$).

The first result, appeared in [19] and refined in [20], treats the rank-one case and is given below.

**Theorem 3.4** ([20, Theorem 6]). *Let $\mathbf{p}_0 = (p_{0,1}, p_{0,2}, \ldots, p_{0,n})$ be a complex-valued vector given as*

$$p_{0,k} = c\lambda^k, \quad k = 1, \ldots, n,$$

*where $\lambda \in \mathbb{C}$.*

- *If $|\lambda| \leq 1$, the solution of (4), i.e.*

$$p_{0,k} = c\lambda^k, \quad k = n+1, \ldots, n+m$$

  *is also a solution of*

$$\widehat{\mathbf{p}} = \arg\min_{\mathbf{p} \in \mathbb{C}^{(n+m)}} \|\mathcal{S}(\mathbf{p})\|_* \text{ subject to } \mathbf{p}_{(1:n)} = \mathbf{p}_0; \qquad (21)$$

  *in particular, if $|\lambda| < 1$, then the solution of (21) is unique.*

- *If $|\lambda| > 1$, then the unique solution of (21) is given by*

$$p_{0,n+k} = c\frac{\lambda^n}{\overline{\lambda}^{-k}}, \quad k = 1, \ldots, m.$$

Theorem 3.4 is illustrated in Fig. 1. Note that the exponential needs to be decreasing (damped), in order for the nuclear norm (21) to give the same solution for when there is no convex relaxation. A similar result was proved for the rank-$r$ case (where the exponentials should be decreasing sufficiently fast).

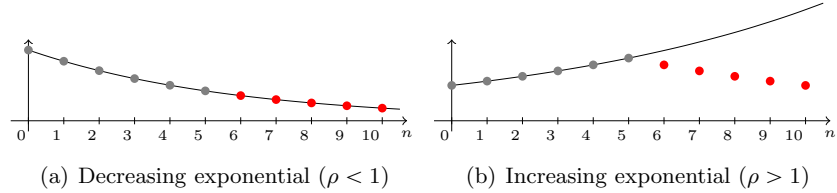

(a) Decreasing exponential ($\rho < 1$)  (b) Increasing exponential ($\rho > 1$)

Figure 1: Forecasts (in red) given by the nuclear norm for the time series $p_k = c\rho^k$, $n = 6$.

**Theorem 3.5** ([20, Theorem 7]). *Fix the structure (20) (i.e. $L = K = n = m + 1$ in (2)) and $r \leq \frac{n}{2}$. Then there exists a number $0 < \rho_{max,r,m} < 1$ such that for any complex-valued time series of finite rank $r$ (14) with*

$$|\lambda_j| < \rho_{max,r,m},$$

*the solution of (21) is unique and coincides with the solution of (18) given in Theorem 3.1.*

Theorem 3.5 implies that the same result holds for the real-valued problems (4) and (5). Note that Theorem 3.5 only requires that the roots $\lambda_j$ of the characteristic polynomial (16) should be all within a disk of radius $\rho_{max,r,m}$ in the complex plane. Theorem 3.5 does not impose any conditions on separation of the roots $\lambda_j$ or their multiplicities $\nu_j$.

**Remark 3.6.** *Theorem 3.4 proves that (in the case of rank-one matrices) the radius $\rho_{max,1,m}$ is equal to 1.*

In the general case ($r > 1$), Theorem 3.5 does not give a good estimate of the radius $\rho_{max,1,m}$.

## 4. Theoretical results on nuclear norm forecasting

### 4.1. Square matrices: fewer missing values

In this subsection, we show that the radius obtained from Theorems 3.4 or 3.5 is also directly applicable to the case of fewer missing values, which is typical in the case of forecasting.

**Corollary 4.1.** *Let $L = K$, $n \geqslant L$, $m = 2L - 1 - n$, and $r < \frac{L+1}{2}$. In this case we have fewer missing values than in Theorem 3.5, i.e. the matrix $\mathcal{S}(\mathbf{p})$ is of the form*

$$\mathcal{S}(\mathbf{p}) = \begin{pmatrix} p_1 & p_2 & \cdots & \cdots & \cdots & p_L \\ p_2 & p_3 & \cdots & \cdot^{\cdot^{\cdot}} & \cdots & \vdots \\ \vdots & \cdots & \cdot^{\cdot^{\cdot}} & \cdot^{\cdot^{\cdot}} & \cdots & p_n \\ \vdots & \cdot^{\cdot^{\cdot}} & \cdot^{\cdot^{\cdot}} & \cdot^{\cdot^{\cdot}} & \cdots & p_{n+1} \\ \vdots & \cdot^{\cdot^{\cdot}} & \cdot^{\cdot^{\cdot}} & \cdot^{\cdot^{\cdot}} & \cdot^{\cdot^{\cdot}} & \vdots \\ p_L & \cdots & p_n & p_{n+1} & \cdots & p_{n+m} \end{pmatrix}. \quad (22)$$

*Let $\rho_{max,r,L}$ be the radius from Theorem 3.5. Then for any complex-valued (or real-valued) time series of the form (14) where*

$$|\lambda_j| < \rho_{max,r,L}$$

*the solution of the nuclear norm minimization (21) for the structure (22) coincides with the solution of the rank minimization problem (18).*

### 4.2. Rectangular matrices: the rank-one case

In this section, we aim to improve the results of the previous subsection. We give an explicit bound for a single (complex) exponential for rectangular matrices.

**Proposition 4.2.** *Let $L, K$ be arbitrary, $m < \min(L, K)$ (so that $n = L + K - 1 - m \geqslant \max(L, K)$). Let $\mathbf{p}_0 = (p_{0,1}, p_{0,2}, \ldots, p_{0,n})$ be a complex-valued time series given as*

$$p_{0,k} = c\lambda^k, \quad (23)$$

*where $\lambda \in \mathbb{C}$. If $\rho = |\lambda|$ satisfies $C(\rho) < 1$, where*

$$C(\rho) = C(\rho, L, K, m) = \begin{cases} |\rho^{m+1} - \rho^{-(m+1)}| \cdot \frac{|\rho|^{L+K}}{\sqrt{|\rho^{2L}-1||\rho^{2K}-1|}}, & |\rho| \neq 1, \\ \frac{m+1}{\sqrt{LK}}, & |\rho| = 1. \end{cases}$$

*then the solution of the nuclear norm minimization (21) for the matrix (2) coincides with the solution of of the rank minimization problem (18).*

As an illustration of Proposition 4.2, Figure 2 contains plots of $C(\rho)$ against $\rho$ for different $m$ with $L = 20$, $K = 50$ and $n = L + K - 1 - m$. Figure 3 contains plots of $C(\rho)$ against $m$ for different $\rho$ with $L = 20$, $K = 50$ and $n = L + K - 1 - m$. In Fig. 3, we can see that for small number of missing values the nuclear norm minimization gives the correct solution if the complex exponential is undamped or increasing in magnitude. Fig. 3 shows that there is for fixed $L, K, \rho$ there is a limiting number of missing data until which the nuclear norm gives the correct solution. In fact, we can find this number explicitly and this is given in Corollary 4.3.

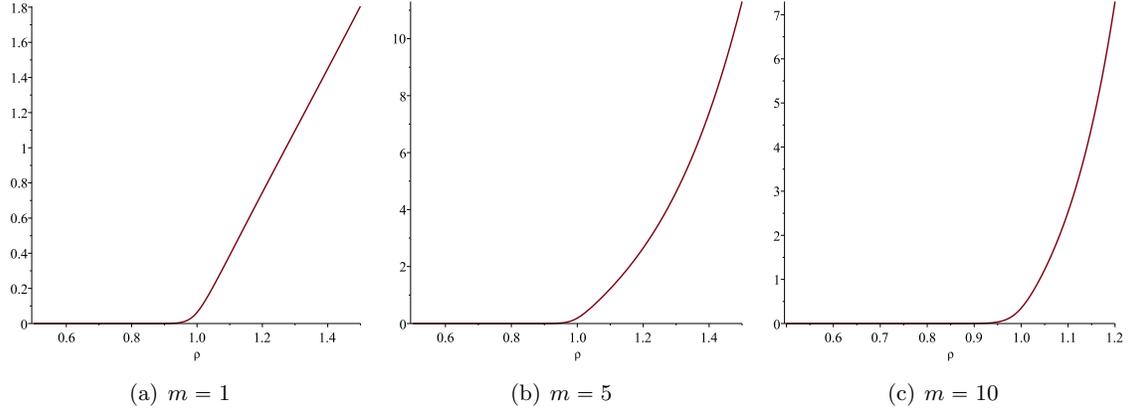

Figure 2: Plot of $C(\rho)$ against $\rho$ for different $m$ with $L = 20$, $K = 50$ and $n = L + K - 1 - m$.

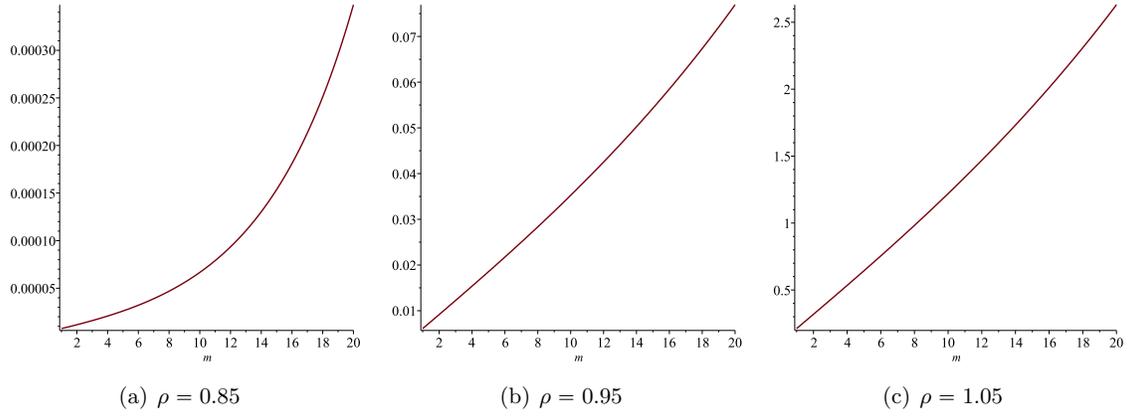

Figure 3: Plot of $C(\rho)$ against $m$ for different $\rho$ with $L = 20$, $K = 50$ and $n = L + K - 1 - m$.

**Corollary 4.3.** *Let $L, K$ and $\rho$ be fixed, and the time series is given as in (23). Then the solution of the nuclear norm minimization (21) for the matrix (2) coincides with the solution of (18) if*

$$m + 1 < \begin{cases} \log_{|\rho|}\left(\frac{\sqrt{y^2+4}-y}{2}\right), & |\rho| < 1, \\ \sqrt{LK}, & |\rho| = 1, \\ \log_{|\rho|}\left(\frac{\sqrt{y^2+4}+y}{2}\right), & |\rho| > 1, \end{cases}$$

*where for $|\rho| \neq 1$ we define*

$$y = \sqrt{\left|1 - \frac{1}{\rho^{2L}}\right|\left|1 - \frac{1}{\rho^{2K}}\right|}.$$

Next we look at the question of choosing the optimal window length. The next corollary shows that the performance of the nuclear norm is maximised when the Hankel matrix is square or almost square.

**Corollary 4.4.** *If $m$ and $n$ are fixed, then $C(\rho)$ is minimised if $L = K$ (in the case $m + n$ odd) or $|L - K| = 1$ (in the case $m + n$ even).*

In Figure 4 we plot $C(\rho)$ against $m$ for different $\rho$ and $L$, which confirms the conclusions of Corollary 4.4. Indeed, the curves for the approximately square shape of the Hankel matrix are lower than the others in Figure 4.

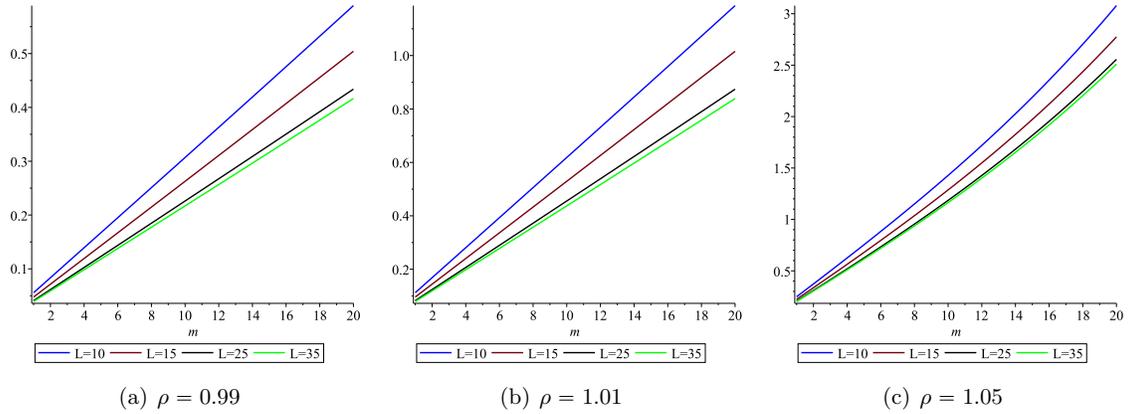

(a) $\rho = 0.99$  (b) $\rho = 1.01$  (c) $\rho = 1.05$

Figure 4: Plot of $C(\rho)$ against $m$ for different $\rho$ and $L$. Here $K = 70 - L$ and $n = L + K - 1 - m$.

Next, we would like to see whether the bound given in Corollary 4.3 is tight. We take $L = K = 13$, the time series $p_k = \rho^k$ and test the performance of the nuclear norm for different values of $\rho$ and $m$. If the Frobenius norm between the minimal rank completion and minimal nuclear norm completion is $\leqslant 10^{-4}$, the nuclear norm completion is declared successful. The results are plotted in Figure 5. From Figure 5 we see that the bound given by Corollary 4.3 is not optimal for small

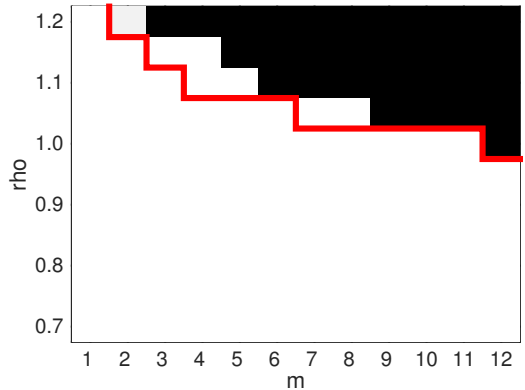

Figure 5: Success of the nuclear norm against $m$ and $\rho$, $L = K = 13$. White: success, black: failure. Red curve: the bound given by Corollary 4.3.

$m$, so the nuclear norm works for higher values of $\rho$.

### 4.3. Rectangular matrices: the rank-r case

First, we provide an improved result from Corollary 4.1 that is valid for rectangular matrices and is stronger so that it may reduce the radius $\rho$.

**Proposition 4.5.** *Let $L$, $K$, $m$, $n$ be as in Theorem 3.1, and consider the matrix structure (2). Fix the number $r \leqslant \frac{\min(L,K)+1}{2}$, and define $m_2 = \max(2r-1, m)$. Let $\rho'_{max,r,m_2}$ be the radius from Theorem 3.5 for matrices of size $(m_2+1) \times (m_2+1)$. Then for any time series of the form (14) where*

$$|\lambda_j| < \rho'_{max,r,m_2}$$

*the solution of the nuclear norm minimization (21) for the full matrix $\mathcal{S}(\mathbf{p})$ coincides with the solution of the rank minimization problem (18) (the same holds for the real-valued problem statements).*

Next, we verify numerically what is the maximum number of missing entries when the nuclear norm gives the correct solution in the rank-$r$ case. For different values of $\rho$, and $r$, we consider the time series

$$p_k = \sum_{j=1}^{r} \rho^j e^{i\omega_j k},$$

where $\omega_j$ are drawn uniformly in $[0, 2\pi)$. For $M = 20$ realisations, we calculate the empirical probability of success, based on the same criterion as in the previous example. The results are plotted in Fig. 6.

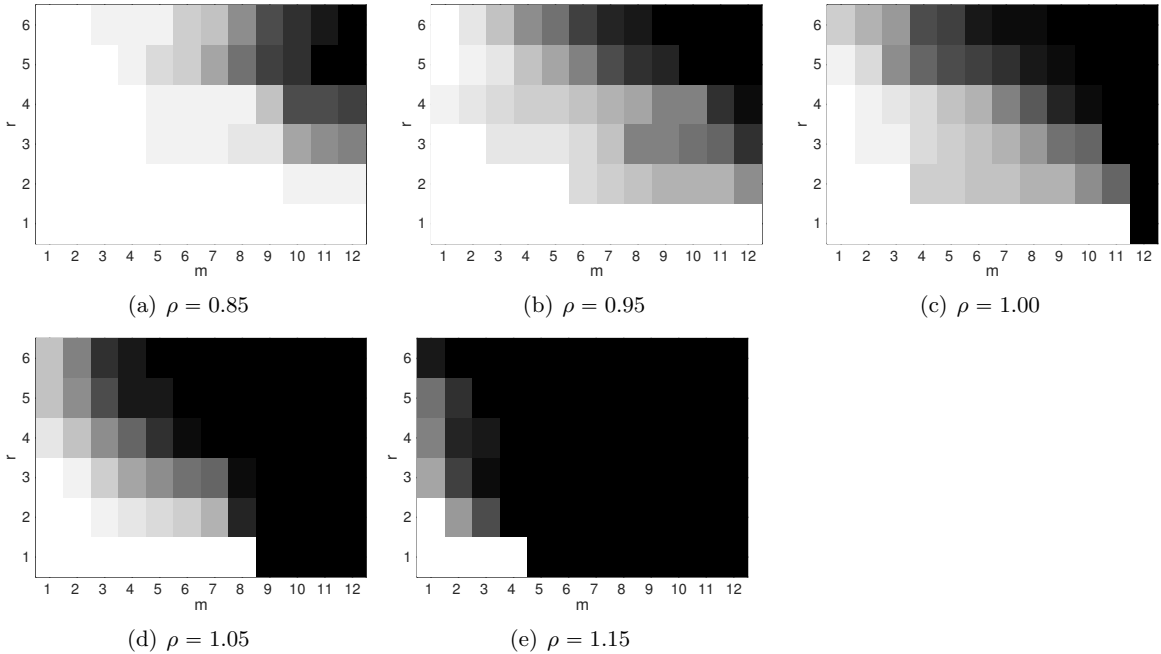

Figure 6: Probability of success of the nuclear norm against $m$, $\rho$ and $r$, $L = K = 13$. Greyscale, white: 1, black: 0.

In Figure 6 we see that the higher the rank is, the smaller number of missing values can be correctly imputed by the nuclear norm. For decaying exponentials, the nuclear norm recovers the correct solution for a large number of missing values and ranks, but for the undamped or increasing exponentials special care has to be taken.

### 4.4. Results and discussion on the approximate matrix completion

The previous subsections suggest that the nuclear norm heuristic works for sufficiently damped sinusoids. Hence, it may be beneficial to preprocess the time series (by introducing an artificial damping) before solving (8). Formally, for a given $\alpha > 0$ we define the scaled matrix structure:

$$\mathcal{S}_{sc}(\mathbf{p}) = \mathcal{S}(\mathbf{p}_{sc}), \tag{24}$$

where

$$\mathbf{p}_{sc} = \left( p_1 \cdot \exp\left(-\frac{\alpha}{2}\right), p_2 \cdot \exp\left(-\frac{2\alpha}{2}\right), \ldots, p_{m+n} \cdot \exp\left(-\frac{(m+n)\alpha}{2}\right) \right) \tag{25}$$

is the scaled time series (multiplied by a decreasing exponential). For simplicity, consider the case of the uniform weights $W_2 = (1, 1, \ldots, 1)$, and solve

$$\mathbf{p}_{***} = \underset{\mathbf{p} \in \mathbb{R}^{(n+m)}}{\arg\min} \, \|\mathcal{S}_{sc}(\mathbf{p})\|_* \text{ subject to } \|\mathbf{p}_{(1:n)} - \mathbf{p}_0\|_{W_2} \leqslant \tau. \tag{26}$$

In fact, we will show in the next proposition that the problem (26) is equivalent to an exponentially weighted problem (8) for the scaled time series.

**Proposition 4.6.** *The solution of problem* (26) *is given by*

1. *Construct the scaled initial data vector* $\mathbf{p}_{sc,0}$ *as in* (25):

$$\mathbf{p}_{sc,0} = \left( p_{0,1} \cdot \exp\left(-\frac{\alpha}{2}\right), p_{0,2} \cdot \exp\left(-\frac{2\alpha}{2}\right), \ldots, p_{0,n} \cdot \exp\left(-\frac{n\alpha}{2}\right) \right).$$

2. *Solve the problem*

$$\mathbf{p}_{sc,*} = \underset{\mathbf{p} \in \mathbb{R}^{(n+m)}}{\arg\min} \, \|\mathcal{S}(\mathbf{p}_{sc})\|_* \text{ subject to } \|\mathbf{p}_{sc,(1:n)} - \mathbf{p}_{sc,0}\|_{W_3} \leqslant \tau, \tag{27}$$

*where* $W_3 = (w_{3,1}, w_{3,2}, \ldots, w_{3,n})$ *such that* $w_{3,i} = \exp(\alpha i), i = 1, \ldots, n$.

3. *Scale back the weighted approximation*

$$\mathbf{p}_{***} = \left( p_{sc,*,1} \cdot \exp\left(\frac{\alpha}{2}\right), p_{sc,*,2} \cdot \exp\left(\frac{2\alpha}{2}\right), \ldots, p_{sc,*,m+n} \cdot \exp\left(\frac{(m+n)\alpha}{2}\right) \right). \tag{28}$$

Hence, by Proposition 4.6 the exponential weighting may help to overcome the potential problems with the increasing exponentials in the time series.

## 5. Examples

### 5.1. Fortified wine

In this section we consider the classical time series 'Fortified Wine', where 120 observations depict the monthly volumes of wine sales in Australia (thousands of litres) in the period from January 1980 until December 1989. Denote by $\mathbf{p}_0$ the vector of these 120 observations, that is, $n = 120$. We do not consider a forecast and thus take $m = 0$.

Figure 7 contains a plot of the vector $\mathbf{p}_0$ with three approximations. The approximations are obtained from (8) with $L = 60$ and weighting scheme (11), so that $\|\mathbf{p} - \mathbf{p}_0\|_{W_1} = \|\mathcal{S}(\mathbf{p}) - \mathcal{S}(\mathbf{p}_0)\|_F$. Define

$$\tau^{(r)} = \min_{X \in \mathbb{R}^{L \times K}, \, \text{rank}(X) = r} \|\mathcal{S}(\mathbf{p}_0) - X\|_F. \qquad (29)$$

For given $r$, the value $\tau^{(r)}$ in (29) can be found using the singular value decomposition, see for example [29, Sect. 2.4]. The motivation for taking $\tau^{(r)}$ as in (29) is to ensure that our approximation $\mathbf{p}_*$ is as 'least as good' as the what would be obtained from the unstructured low rank approximation of $\mathcal{S}(\mathbf{p}_0)$. We consider three values of $\tau^{(r)}$, with $r = 1, 3$ and $10$.

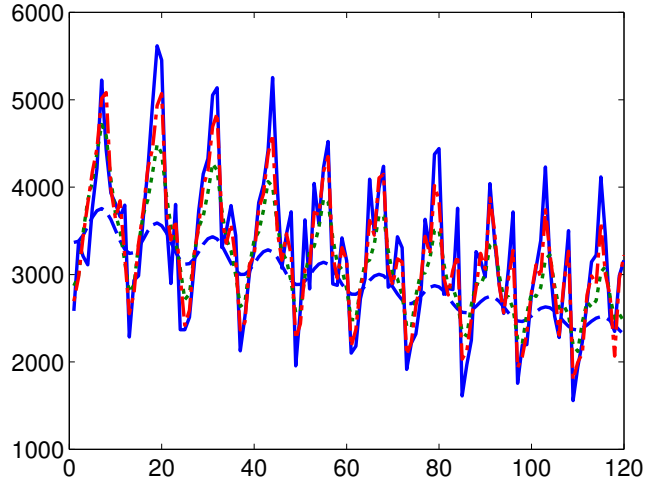

Figure 7: Fortified wine (solid line) with three approximations obtained from (8), $\tau = \tau^{(1)}$ (long dash), $\tau = \tau^{(3)}$ (short dash) and $\tau = \tau^{(10)}$ (dot-dash).

Figure 8 contains plots of the square root of the first fifteen singular values of the matrix $\mathcal{S}(\mathbf{p}_*)$ where $\mathbf{p}_*$ is the solution to (8), for three values of $\tau$ as described above. Minimizing the sum of singular values results in many individual singular values going close to zero. The parameter $\tau$ can be used to control the complexity of the approximation, the smaller the value of $\tau$, the closer the approximation will be to the given vector $\mathbf{p}_0$.

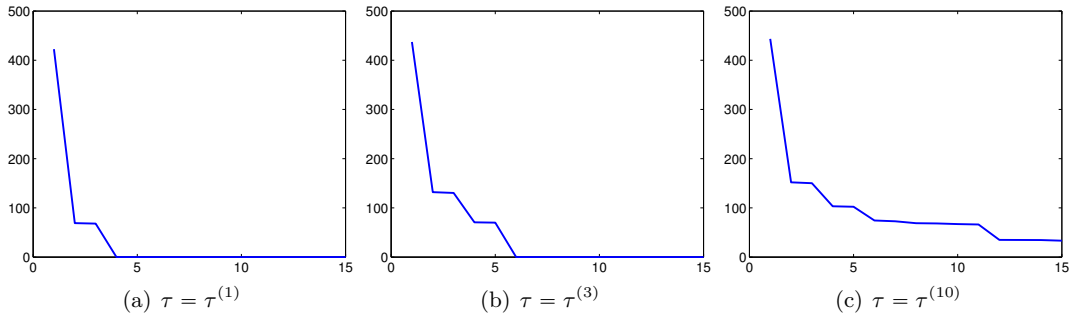

(a) $\tau = \tau^{(1)}$  (b) $\tau = \tau^{(3)}$  (c) $\tau = \tau^{(10)}$

Figure 8: Plot of the square root of the first 15 singular values of the matrix $\mathcal{S}(\mathbf{p}_*)$ where $\mathbf{p}_*$ is the solution to (8).

### 5.2. Forecasting deaths

In this section we consider forecasting the famous 'death' series recording the monthly accidental deaths in the USA between 1973 and 1978. This data has been studied by many authors (such as [2]) and can be found in a number of time series data libraries. We wish to replicate the exercise given in [30] which aimed to forecast the final six values of this series. The time series contains a total of 78 observations. We truncate the series to the first 72 observations and will forecast the remaining six observations. Denote these series of 72 observations by $\mathbf{p}_0$. We consider (8) with $n = 72$ and $m = 6$.

Table 1 contains forecasts of the final six data points of the data series by several methods along with the square root of the mean square error. These results are taken from [30] and full details of the fitted models can be found within. In summary Model I and Model II are examples of SARIMA models as described by [31]. Model I is given by

$$\nabla_{12} p_i = 28.831 + (1 - 0.478B)(1 - 0.588B^{12})Z_i$$

and Model II is given by

$$\nabla_{12} p_i = 28.831 + Z_i - 0.596 Z_{i-1} - 0.407 Z_{i-6} - 0.685 Z_{i-12} + 0.460 Z_{i-13}$$

where $Z_i$ is a realisation of white noise with zero mean and variance 0.9439 and $B$ is the backward shift operator defined as: $B^j Z_i = Z_{i-j}$. HWS represents the model as fitted by the Holt-Winter seasonal algorithm. ARAR represents the model as fitted by transforming the data prior to fitting an autoregressive model.

|  | 1 | 2 | 3 | 4 | 5 | 6 | $\sqrt{MSE}$ |
|---|---|---|---|---|---|---|---|
| Original data | 7798 | 7406 | 8363 | 8460 | 9217 | 9316 |  |
| Model I | 8441 | 7704 | 8549 | 8885 | 9843 | 10279 | 582.626 |
| Model II | 8345 | 7619 | 8356 | 8742 | 9795 | 10179 | 500.500 |
| HWS | 8039 | 7077 | 7750 | 7941 | 8824 | 9329 | 401.263 |
| ARAR | 8168 | 7196 | 7982 | 8284 | 9144 | 9465 | 253.202 |

Table 1: Forecasted data using four different models, along with the square root of the mean square error.

We now consider the following exercise. We take $L = 24$ and consider three forms of $W$ defined in Section 2.3. For the weighting scheme $W_3$ we take $\alpha = 0.05$. We will solve (8) under these three weighting schemes and select $\tau$ such that the solutions obtained are as least as close to the solutions obtained by unstructured rank 3, 6 and 12 approximations to $\mathcal{H}_L(\mathbf{p}_0)$. More precicely, for a weight vector $W$ and rank $r$, we take $\tau$ to be the solution of

$$\check{X} = \arg\min_{X \in \mathbb{R}^{L \times K},\, \text{rank}(X) = r} \|\mathcal{H}_L(\mathbf{p}_0) - X\|_F,$$

$$\check{\mathbf{p}} = \arg\min_{\mathbf{p} \in \mathbb{R}^n} \|\mathcal{H}_L(\mathbf{p}) - \check{X}\|_F,$$

$$\tau^{(r)} = \|\check{\mathbf{p}} - \mathbf{p}_0\|_W.$$

In particular, for the weighting scheme $W_1$ it is equivalent to taking $\tau = \tau^{(3)}, \tau^{(6)}$ and $\tau^{(12)}$ as in (29). Also, note that $\check{\mathbf{p}}$ can be obtained by orthogonal projection on the space of the Hankel matrices. Hence, it can be computed by diagonal averaging [2, Sec. 6.2], and is, in fact, the SSA approximation of $\mathbf{p}_0$. The results are given in Table 2.

|  | 1 | 2 | 3 | 4 | 5 | 6 | $\sqrt{MSE}$ |
|---|---|---|---|---|---|---|---|
| Original data | 7798 | 7406 | 8363 | 8460 | 9217 | 9316 |  |
| $W_1$, rank = 3 | 7829 | 7730 | 7775 | 8045 | 8485 | 8815 | 485.06 |
| $W_1$, rank = 6 | 7860 | 7722 | 7779 | 8030 | 8714 | 9006 | 404.00 |
| $W_1$, rank = 12 | 7873 | 7648 | 7808 | 8076 | 8868 | 9123 | 336.27 |
| $W_2$, rank = 3 | 8371 | 8219 | 8233 | 8501 | 8945 | 9356 | 424.87 |
| $W_2$, rank = 6 | 8269 | 8007 | 8022 | 8262 | 9081 | 9401 | 356.75 |
| $W_2$, rank = 12 | 8096 | 7704 | 7916 | 8123 | 9040 | 9293 | 295.23 |
| $W_3$, rank = 3 | 8233 | 7947 | 7863 | 8236 | 9485 | 10012 | 472.29 |
| $W_3$, rank = 6 | 8029 | 7444 | 7874 | 8349 | 9433 | 9782 | 308.28 |
| $W_3$, rank = 12 | 7953 | 7490 | 7906 | 8118 | 9122 | 9356 | 247.38 |

Table 2: Forecasted data using nine different models, along with the square root of the mean square error.

Figure 9 contains plots of $\mathbf{p}_0$ along with approximations $\mathbf{p}_0$ obtained from (8) with $W$ and $\tau$ as defined above. The effects of the different weighting schemes can clearly be seen, also the effect of decreasing $\tau$ is visible.

Figure 10 contains a plot of the logarithm of the square root of the mean square error of the forecast against $\alpha$ and $\tau$ for weighting scheme $W_3$. The smallest square root of the mean square error is 219.91 obtained at $\alpha = 0.01$ and $\tau = 8000$.

We now make some remarks:

- As the rank increases, or more formally, as $\tau$ increases, then the quality of the forecast improves.

- $W_1$ appears to be the worst weighting scheme. Recall that in this case the weights are given in (11). It can be seen that this weighting has the unfortunate characteristic of 'down-weighting' observations towards the end of the vector $\mathbf{p}_0$, contradictory to the argument that the more recent observations are more important for forecasting.

- Weighting scheme $W_3$ gives increased weight to the more recent observations. It can be seen that this weighting scheme gives the best forecast, in the sense that it provides a forecast with the smallest mean square error (and smaller than the best mean square error found in [30]).

5.3. Simulation study

We now consider a simulated example where we consider a time series $\mathbf{p}_0$ of $n = 100$ observations of the form $p_i = s_i + \varepsilon_i$ where $s_i = \cos(2\pi i/10)$ (denoted Case 1) or $s_i = \exp(0.02i)\cos(2\pi i/10)$ (denoted Case 2) and $\varepsilon_i$ is a white noise (Gaussian) error term with zero mean and standard deviation $\sigma = 0.1$. We truncate the series to the first $n - m$ terms and the aim of this study is to forecast the remaining $m$ observations for $m = 1, 2, \ldots, 15$. We take $L = 30$ in this example.

Figures 11 and 12 contain the square root of the mean square errors obtained from forecasting the $m$ remaining observations, using (8) under two possible weighting schemes, for both Case 1 and Case 2. We consider three weighting schemes ($W_1, W_2, W_3$) defined in Section 2.3. We select $\tau$ such that the solutions obtained are as least as close to the unstructured rank 2 approximation $\mathcal{S}(\mathbf{p}_0)$ in each of the weighting schemes given above, in the same way as in subsection 5.2.

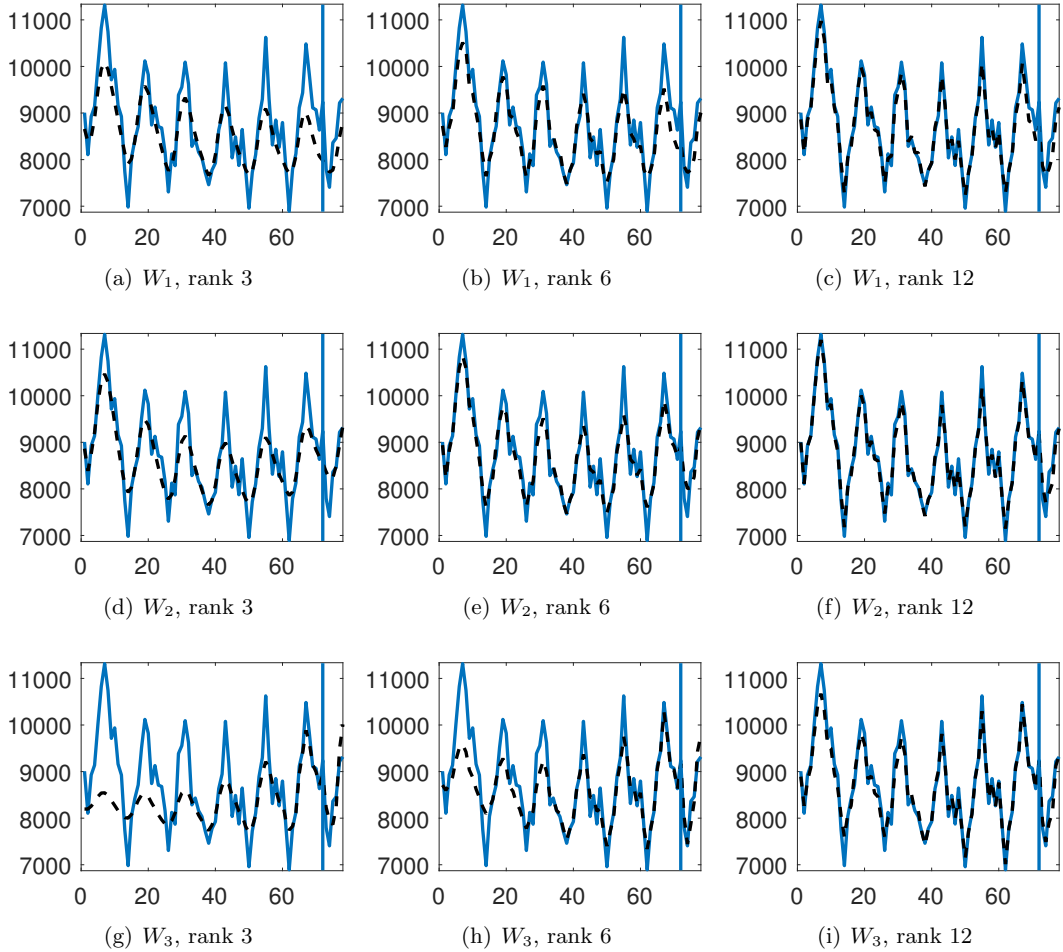

Figure 9: Plot of the data (solid line) with approximation and forecasts (dashed line).

We remark that weighting scheme $W_3$ provides the better forecasts, especially for Case 2. Given the form of the time series in Case 2, this result is to be expected. The time series in this case grows exponentially in time. This gives an indication as to the merits of alternative weighting schemes, as opposed to the 'traditional' weighting scheme given by $W_1$.

## 6. Conclusion

In this paper we have considered matrix completion as a tool for forecasting in time series analysis. We have formulated a nuclear norm relaxation of structured low-rank matrix completion suitable for this purpose, and have demonstrated its practical potential towards the end of the paper. We have shown that the time series should be sufficiently damped (i.e. the exponentials should not be increasing very fast) for the nuclear norm approach to work. This becomes par-ticularly important when the rank of the time series or the number of values to be forecasted is large. An approach based on exponential weighting is proposed, which is shown to be equivalent to preprocessing of time series. Numerical experiments indicate that the use of exponential weighting improves the performance of the nuclear norm forecasting.

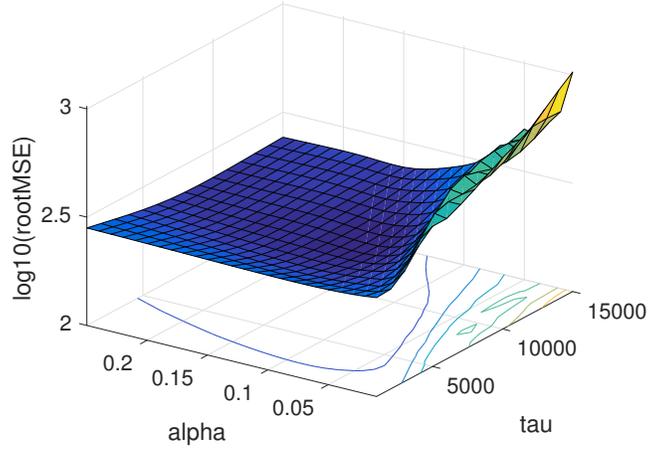

Figure 10: Logarithm of the square root of the mean square error of the forecast against $\alpha$ and $\tau$ for weighting scheme $W_3$.

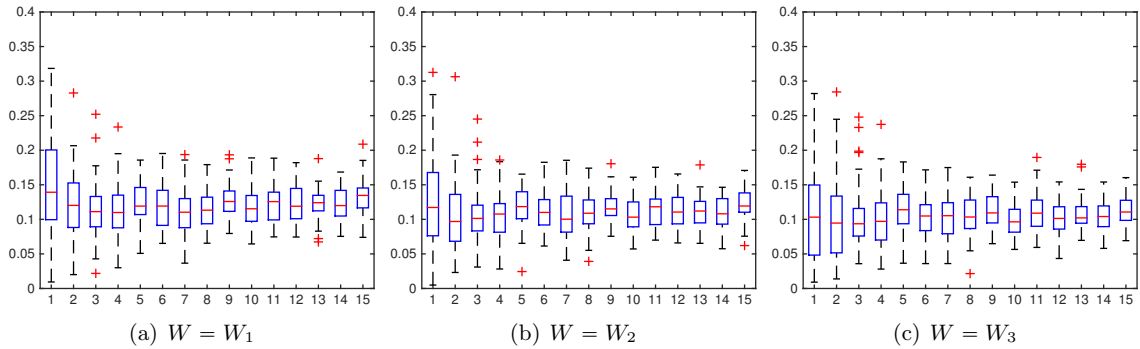

Figure 11: Case 1; Square root of the mean square errors against the number of forecasted observations $m$ for three weighting schemes. Taken over 50 simulations.

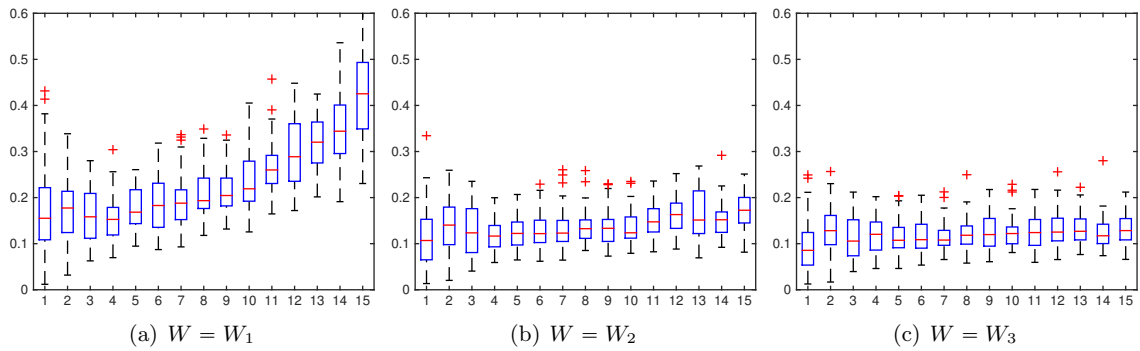

Figure 12: Case 2; Square root of the mean square errors against the number of forecasted observations $m$ for three weighting schemes. Taken over 50 simulations.

## Appendix A. Structured matrices and optimality conditions

The proofs of Theorems 3.5 and 3.4 and are based on the following necessary and sufficient condition for a global optimum of (5) or (21).

**Lemma Appendix A.1** (A generalization of [20, Proposition 14]). *Let $\widehat{\mathbf{p}}_* \in \mathbb{C}^{n+m}$ and the compact SVD of $\mathcal{S}(\widehat{\mathbf{p}}_*)$ be given by*
$$\mathbf{U}\mathbf{\Sigma}\mathbf{V}^H = \mathcal{S}(\widehat{\mathbf{p}}_*).$$
*Further, let $\mathbf{Q}_1 = \mathbf{U}\mathbf{U}^H$ and $\mathbf{Q}_2 = \mathbf{V}\mathbf{V}^H$ and $\mathbf{B} = \mathbf{U}\mathbf{V}^H$. Then the following statements hold true.*

1. *$\widehat{\mathbf{p}}_*$ is a global minimizer of (21) (and a minimizer of (5) is $\mathcal{S}(\widehat{\mathbf{p}}_*)$ is real) if and only if there exists a matrix $\mathbf{M} \in \mathbb{R}^{m \times n}$ such that $\|\mathbf{M}\|_2 \leq 1$ and*

$$\langle \mathbf{Q}_1 \mathbf{M} \mathbf{Q}_2 + \mathbf{B}, \mathbf{S}_k \rangle = 0, \tag{A.1}$$

   *holds for all $k \in \{n+1, \ldots, n+m\}$.*
2. *If, in addition, $\|\mathbf{M}\|_2 < 1$, then $\widehat{\mathbf{p}}_*$ is the unique minimizer of (21) (and the minimizer (5) is $\mathcal{S}(\widehat{\mathbf{p}}_*)$ is real).*

The matrix $\mathbf{M}$ in Lemma Appendix A.1 is called the optimality certificate. In fact, the proofs of Theorems 3.5 and 3.4 proceed by constructing such an optimality certificate.

## Appendix B. Proofs

*Proof of Theorem 3.1.* The proof follows from [26, Theorem 5.13], applied to a submatrix of $\mathcal{S}(\mathbf{p})$. More precisely, under the conditions of the theorem, the matrix $\mathcal{S}(\mathbf{p})$ contains an $r \times (r+1)$ matrix of rank $r$ with only known entries (no missing values). By [26, §5] and [26, Theorem 5.13] in particular, there exists only one extension to an $L \times K$ matrix that preserves the rank $r$ (any other extension would increase the rank). This is exactly the completion given in (19). □

*Proof of Corollary 4.1.* Let $\tilde{\mathbf{p}}$ be the solution given in (19). By Theorem 3.5 the solution of (18) is the unique minimiser of (21) for the matrix (20). Moreover, from the proof of Theorem 3.5 (see also Lemma Appendix A.1 and the remark after it), there exists a matrix $\mathbf{M} \in \mathbb{R}^{L \times L}$, $\|\mathbf{M}\|_2 < 1$ such equation (A.1) holds for all $k \in \{L+1, \ldots, n+m\}$. Hence, in particular, equation (A.1) holds for all $k \in \{n+1, \ldots, n+m\}$, and by Lemma Appendix A.1, the point $\tilde{\mathbf{p}}$ is also the unique minimiser of (21) for the structure as given in (22). □

*Proof of Proposition 4.2.* The solution of (4) is given by continuing the formula (23), i.e. $\tilde{p}_k = c\lambda^k$. Let us show that the point $\tilde{\mathbf{p}}$ is optimal for (18).

Denote the matrix $\mathbf{X} = S(\tilde{\mathbf{p}})$. Since the matrix is rank-one, its compact SVD has the form $\mathbf{X} = \sigma \mathbf{u}\mathbf{v}^H$, and therefore, we have that $\mathbf{B} = \frac{\mathbf{X}}{\sigma}$.

Let us find explicitly the terms of the compact SVD. The matrix $\mathbf{X}$ is rank-one, so $\mathbf{X} = c\mathbf{a}\mathbf{b}^H$, where

$$\mathbf{a} = \begin{pmatrix} 1 \\ \lambda \\ \vdots \\ \lambda^{L-1} \end{pmatrix}, \quad \mathbf{b} = \begin{pmatrix} 1 \\ \overline{\lambda} \\ \vdots \\ \overline{\lambda}^{K-1} \end{pmatrix}.$$

Assume that $\rho = |\lambda|$, $\lambda = \rho \exp(\jmath\omega)$, $\jmath$ is the imaginary unit, and $c = |c|\exp(\jmath\phi)$. Then the singular vectors $\mathbf{u}$, $\mathbf{v}$ of a compact SVD can be found as

$$\mathbf{u} = \frac{\mathbf{a}\exp(\jmath\phi)}{\|\mathbf{a}\|_2}, \quad , \mathbf{v} = \frac{\mathbf{b}}{\|\mathbf{b}\|_2}.$$

hence the singular value $\sigma$ can be expressed as

$$\sigma = |c|\|\mathbf{a}\|_2\|\mathbf{b}\|_2 = |c| \cdot \begin{cases} \frac{\sqrt{|(\rho^{2L}-1)(\rho^{2K}-1)|}}{|\rho^2-1|}, & |\rho| \neq 1, \\ \sqrt{LK}, & |\rho| = 1. \end{cases}$$

In addition, the matrices $\mathbf{B}$, $\mathbf{Q}_1$ and $\mathbf{Q}_2$ defined in Lemma Appendix A.1 are given by

$$\mathbf{Q}_1 = \left(\mathbf{I} - \frac{\mathbf{a}\mathbf{a}^H}{\|\mathbf{a}\|_2^2}\right), \quad \mathbf{Q}_2 = \left(\mathbf{I} - \frac{\mathbf{b}\mathbf{b}^H}{\|\mathbf{b}\|_2^2}\right).$$

In what follows, we use the optimality condition given in Lemma Appendix A.1. Let us find a matrix $\mathbf{M}$ that satisfies (A.1). We will do that by truncating the matrix $\mathbf{X}$ and using [20, Theorem 6] for the truncated matrix.

Consider the lower right $(m+1) \times (m+1)$ submatrix of the matrix in (22)

$$\mathcal{S}'(\mathbf{p}) = \begin{pmatrix} p_{n-m} & p_{n-m+1} & \cdots & p_n \\ p_{n-m+1} & p_{n-m+2} & \cdot\cdot\cdot & p_{n+1} \\ \vdots & \cdot\cdot\cdot & \cdot\cdot\cdot & \vdots \\ p_n & p_{n+1} & \cdots & p_{n+m} \end{pmatrix}. \tag{B.1}$$

Then the matrix $\mathbf{X}' = \mathcal{S}'(\tilde{\mathbf{p}})$ has the form

$$\mathbf{X}' = c\lambda^{n-m}\mathbf{y}\mathbf{y}^T, \quad \mathbf{y} = \begin{pmatrix} 1 \\ \lambda \\ \vdots \\ \lambda^m \end{pmatrix},$$

and the singular vectors in its compact SVD $\mathbf{X}' = \sigma'\mathbf{u}'\mathbf{v}'^H$ can be chosen as

$$\mathbf{u}' = \frac{\mathbf{y}\exp(\jmath(\phi + \omega(m-n)))}{\|\mathbf{y}\|_2}, \quad \mathbf{v} = \frac{\mathbf{y}}{\|\mathbf{y}\|_2}.$$

Hence, the singular value is equal to

$$\sigma' = |c||\rho|^{n-m}\|\mathbf{y}\|_2^2 = |c||\rho|^{n-m} \cdot \begin{cases} \frac{|\rho^{2(m+1)}-1|}{|\rho^2-1|}, & |\rho| \neq 1, \\ m+1, & |\rho| = 1. \end{cases}$$

Again, we denote $\mathbf{B}' = \mathbf{u}'\mathbf{v}'^H$, $\mathbf{Q}' = (\mathbf{I} - \mathbf{u}'\mathbf{u}'^H)$, where the matrix $\mathbf{B}'$ can be found as $\mathbf{B}' = \frac{\mathbf{X}'}{\sigma'}$.

Let us construct the optimality certificate. By Theorem 3.4 ([20, Theorem 6]), there exists a matrix $\mathbf{M} \in \mathbb{C}^{(m+1) \times (m+1)}$ such that

$$\langle \mathbf{B}' + \mathbf{Q}'\mathbf{M}'\mathbf{Q}'^T, \mathbf{S}'_k \rangle = 0, \quad \forall k \in \{n+1, \ldots, (n+m)\}.$$

and also $\|\mathbf{M}'\|_2 = |\rho|^m$, $\mathbf{Q}'\mathbf{M}'\mathbf{Q}'^T = \mathbf{M}'$.

Next, we define the matrix $\mathbf{M} \in \mathbb{C}^{L \times K}$ as follows:

$$\mathbf{M} = \frac{\sigma'}{\sigma} \begin{pmatrix} 0 & 0 \\ 0 & \mathbf{M}' \end{pmatrix}.$$

Then since $\mathbf{Q}\mathbf{M}\mathbf{Q}^T = \mathbf{M}$ we have that

$$\langle \mathbf{B} + \mathbf{Q}\mathbf{M}\mathbf{Q}^T, \mathbf{S}_k \rangle = \langle \mathbf{B}, \mathbf{S}_k \rangle + \langle \mathbf{M}, \mathbf{S}_k \rangle,$$

where the first term can be found as

$$\langle \mathbf{B}, \mathbf{S}_k \rangle = \frac{1}{\sigma}\langle \mathbf{X}, \mathbf{S}_k \rangle = \frac{1}{\sigma}\langle \mathbf{X}', \mathbf{S}'_k \rangle = \frac{\sigma'}{\sigma}\langle \mathbf{B}', \mathbf{S}'_k \rangle$$

Hence we have that

$$\langle \mathbf{B} + \mathbf{Q}\mathbf{M}\mathbf{Q}^T, \mathbf{S}_k \rangle = \frac{\sigma'}{\sigma}\langle \mathbf{B}', \mathbf{S}'_k \rangle + \frac{\sigma'}{\sigma}\langle \mathbf{M}', \mathbf{S}'_k \rangle = 0,$$

and $\mathbf{M}$ satisfies (A.1). Since, by construction, $\|\mathbf{M}\|_2 = \frac{\sigma'}{\sigma}|\rho|^m = C(\rho)$, the proof is complete by Lemma Appendix A.1. □

*Proof of Corollary 4.4.* For fixed $m, n$, the value of $C(\rho)$ is minimized if

$$\begin{cases} \sqrt{LK}, & |\rho| = 1, \\ \sqrt{\left|1 - \frac{1}{\rho^{2L}}\right|\left|1 - \frac{1}{\rho^{2K}}\right|}, & |\rho| \neq 1, \end{cases}$$

is maximized. It is clear that this quantity is a geometric mean of two quantities that monotonically depend on $L$ and $K$, which is maximised if $L$ and $K$ are as close as possible to each other. □

Before proving Proposition 4.5, we will need a technical lemma about the bases of subspaces.

**Lemma Appendix B.1.** *Let $\mathbf{A} = \begin{pmatrix} \mathbf{A}^{(1)} \\ \mathbf{A}^{(2)} \end{pmatrix} \in \mathbb{C}^{n \times r}$, where $\mathbf{A}^{(1)} \in \mathbb{C}^{n_1 \times r}$, $\mathbf{A}^{(2)} \in \mathbb{C}^{n_2 \times r}$ such that $\operatorname{rank} \mathbf{A} = \operatorname{rank} \mathbf{A}^{(2)} = r$. Let $\mathbf{U}^{(2)} \in \mathbb{C}^{n_2 \times r}$ and $\mathbf{U} \in \mathbb{C}^{n \times r}$ be the orthonormal bases for column spaces of $\mathbf{A}^{(2)}$ and $\mathbf{A}$ respectively.*

*Further, let $m \leqslant n_2$ and $\mathbf{U}_{\downarrow m}$, $\mathbf{U}^{(2)}_{\downarrow m}$ be the matrices containing last $m$ rows of $\mathbf{U}$, $\mathbf{U}^{(2)}$ respectively:*

$$\mathbf{U}_{\downarrow m} = \mathbf{U}_{n-m+1:n,:}, \quad \mathbf{U}^{(2)}_{\downarrow m} = \mathbf{U}^{(2)}_{n_2-m+1:n_2,:}.$$

*Then it holds that*

$$\|\mathbf{U}_{\downarrow m}\|_F \leqslant \|\mathbf{U}^{(2)}_{\downarrow m}\|_F. \tag{B.2}$$

*Proof.* First, we only need to prove (B.2) for $n_1 = 1$, so that $\mathbf{A}^{(2)}$ is $\mathbf{A}$ with the first row removed. The case $n_1 > 1$ can be proved by recursively removing the first row $n_1$ times. Second, the quantities in (B.2) do not depend on particular bases of subspaces due to invariance of the Frobenius norm under the unitary transformations. Hence, we can assume, for example, that $\mathbf{U}$ and $\mathbf{U}^{(2)}$ are obtained from the thin QR factorizations, i.e. $\mathbf{A} = \mathbf{U}\mathbf{R}$ and $\mathbf{A}^{(2)} = \mathbf{U}^{(2)}\mathbf{R}^{(2)}$.

Consider the case $n_1 = 1$, $n_2 = n - 1$, and the full QR factorization of $\mathbf{A}$:

$$\mathbf{A} = \mathbf{Q}\begin{pmatrix}\mathbf{R}\\ \mathbf{0}\end{pmatrix}, \quad \text{where } \mathbf{Q} = \begin{pmatrix}\mathbf{U} & \mathbf{W}\end{pmatrix}, \quad \mathbf{W} \in \mathbb{C}^{n \times (n-r)}.$$

Without loss of generality, we can assume that $\mathbf{W}$ has the form

$$\mathbf{W} = \begin{pmatrix}\beta & \mathbf{0}\\ \mathbf{w} & *\end{pmatrix},$$

where $\mathbf{w} \in \mathbb{C}^{n_2}$ and $\beta \in \mathbb{C}$, i.e. only one element in the first row of $\mathbf{W}$ is nonzero. Next, we use the fact that $\mathbf{A}^{(2)}$ is obtained from $\mathbf{A}$ by removing the first row. From [29, Sec. 12.5.3] and from the structure of $\mathbf{W}$ we have that

$$\mathbf{A} = \begin{pmatrix}\mathbf{A}^{(1)}\\ \mathbf{A}^{(2)}\end{pmatrix} = (\mathbf{Q}\mathbf{G}_r \cdots \mathbf{G}_1)\begin{pmatrix}\mathbf{v}^T\\ \mathbf{R}^{(2)}\\ \mathbf{0}\end{pmatrix} = \begin{pmatrix}\pm 1 & \mathbf{0}\\ \mathbf{0} & \mathbf{Q}^{(2)}\end{pmatrix}\begin{pmatrix}\mathbf{v}^T\\ \mathbf{R}^{(2)}\\ \mathbf{0}\end{pmatrix} \quad (B.3)$$

where $\mathbf{v}$ is a vector and $\mathbf{G}_k \in \mathbb{C}^{n \times n}$, $k \in \{1,\ldots,r\}$ are appropriate Givens rotations rotating in the plane of the indices $(k, k+1)$, and $\mathbf{Q}^{(2)} = \begin{pmatrix}\mathbf{U}^{(2)} & \mathbf{W}^{(2)}\end{pmatrix}$, so that $\mathbf{A}^{(2)} = \mathbf{Q}^{(2)}\begin{pmatrix}\mathbf{R}^{(2)}\\ \mathbf{0}\end{pmatrix}$ is the full QR factorization of $\mathbf{A}^{(2)}$. Finally, since the rotation matrices $\mathbf{G}_r, \ldots, \mathbf{G}_1$ affect only first $r+1$ columns of $\mathbf{Q}$, we have that

$$\begin{pmatrix}\mathbf{U} & \begin{pmatrix}\beta\\ \mathbf{w}\end{pmatrix}\end{pmatrix}\widetilde{\mathbf{G}}_r \cdots \widetilde{\mathbf{G}}_1 = \begin{pmatrix}\pm 1 & \mathbf{0}\\ \mathbf{0} & \mathbf{U}^{(2)}\end{pmatrix}$$

where $\widetilde{\mathbf{G}}_r \in \mathbb{C}^{(r+1) \times (r+1)}$ are the reduced Givens rotation matrices. Hence, if $\mathbf{w}_{\downarrow m}$ is the vector of the last $m$ elements of $\mathbf{w}$, then

$$\|\mathbf{U}^{(2)}_{\downarrow m}\|_F^2 = \|\mathbf{U}_{\downarrow m}\|_F^2 + \|\mathbf{w}_{\downarrow m}\|_2^2 \leqslant \|\mathbf{U}_{\downarrow m}\|_F^2,$$

which completes the proof. □

*Proof of Proposition 4.5.* We are going to prove that there exists such a (certificate) matrix $\mathbf{M} \in \mathbb{C}^{K \times L}$ that satisfies (A.1). Similarly to the proof for Proposition 4.2, we are going to construct the certificate in the form

$$\mathbf{M} = \begin{pmatrix}0 & 0\\ 0 & \mathbf{M}'\end{pmatrix}, \quad \mathbf{M}' \in \mathbb{C}^{(m_2+1) \times (m_2+1)}.$$

Then (A.1) can be rewritten as where

$$\langle \mathbf{B}_{red} + \mathbf{Q}'\mathbf{M}'\mathbf{Q}'^T, \mathbf{S}'_k\rangle = 0, \quad (B.4)$$

where $\mathbf{B}_{red} \in \mathbb{C}^{(m+1)\times(m+1)}$ is the lower right submatrix of $\mathbf{B}$ and $\mathbf{Q}' = \mathbf{I}_{m_2+1} - \mathbf{P}' \in \mathbb{C}^{(m_2+1)\times(m_2+1)}$ and $\mathbf{P}'$ the projector on the column space of the lower right $(m_2+1)\times(m_2+1)$ submatrix of $\mathcal{S}(\mathbf{p})$ defined as

$$\mathcal{S}'(\mathbf{p}) = \begin{pmatrix} p_{n+m-2m_2} & p_{n+m-m_2+1} & \cdots & \cdots & \cdots & p_{n+m-m_2} \\ p_{n+m-m_2+1} & p_{n+m-m_2+2} & \cdots & \ddots & \cdots & \vdots \\ \vdots & \cdots & \ddots & \ddots & \cdots & p_n \\ \vdots & \ddots & \ddots & \ddots & \cdots & p_{n+1} \\ \vdots & \ddots & \ddots & \ddots & \ddots & \vdots \\ p_{n+m-m_2} & \cdots & p_n & p_{n+1} & \cdots & p_{n+m} \end{pmatrix}. \tag{B.5}$$

We are going to look for the matrix $\mathbf{M}'$ that satisfies (B.4) for $k \in \{n+m-m_2+1, \ldots, n+m\}$ ( a few extra constraints are added). As in [20, eqn. (23)], we rewrite the constraints as

$$\mathcal{A}(\mathbf{P}') \operatorname{vec}(\mathbf{M}') = -(\mathbf{S}')^T \operatorname{vec}(\mathbf{B}_{red}), \tag{B.6}$$

where

$$\mathbf{S}' = \begin{pmatrix} \operatorname{vec}(\mathbf{S}_{n+m-m_2+1}) & \cdots & \operatorname{vec}(\mathbf{S}_{n+m}) \end{pmatrix},$$

and

$$\mathcal{A}(\mathbf{P}') = (\mathbf{S}')^T((\mathbf{I}_{m_2+1} - \mathbf{P}') \otimes (\mathbf{I}_{m_2+1} - \mathbf{P}')) = (\mathbf{S}')^T(\mathbf{Q}' \otimes \mathbf{Q}').$$

are the same matrices as in [20, eqn. (21)–(22)].

Next, as in the proof of [20, Lemma 21] we are going to find $\mathbf{M}'_*$ such that it has the minimal Frobenius norm and satisfies (B.4). Such a matrix is given by

$$\operatorname{vec}(\mathbf{M}'_*) = -\mathcal{A}(\mathbf{P}')^\dagger (\mathbf{S}')^T \operatorname{vec}(\mathbf{B}_{red}) = -\mathcal{A}(\mathbf{P}')^\dagger (\mathbf{S}')^T \operatorname{vec}(\mathbf{B}_{red} - \mathbf{P}'_0 \mathbf{B}_{red}(\mathbf{P}'_0)^T),$$

where, as in [20, eqn. (27)]

$$\mathbf{M} = \begin{pmatrix} \mathbf{I}_r & \mathbf{0} \\ \mathbf{0} & \mathbf{0} \end{pmatrix} \in \mathbb{C}^{(m_2+1)\times(m_2+1)}$$

is the projector on the subspace spanned by the first $r$ unit vectors.

As in [20, eqn. (31)], we note that

$$\|\mathbf{M}'_*\|_2 \leq \|\operatorname{vec}(\mathbf{M}'_*)\|_2 \leq \|\mathcal{A}(\mathbf{P}')^\dagger\|_2 \|(\mathbf{S}')^T\|_2 \|\mathbf{B}_{red} - \mathbf{P}'_0 \mathbf{B}_{red}(\mathbf{P}'_0)^T\|_F.$$

Now if prove that, as in [20, Lemma 20],

$$\|\mathbf{B}_{red} - \mathbf{P}'_0 \mathbf{B}_{red}(\mathbf{P}'_0)^T\|_F \leq \|\mathbf{P}' - \mathbf{P}'_0\|_F.$$

then $\|\mathbf{M}'_*\|_2$ will be bounded by the right hand side of [20, eqn. (31)], and the result will hold true analogously to [20, Theorem 7] (which only relies on [20, eqn. (31)]).

As in the proof of [20, Lemma 20], we have that

$$\begin{aligned} \|\mathbf{B}_{red} - \mathbf{P}'_0 \mathbf{B}_{red}(\mathbf{P}'_0)^T\|_F^2 &\leq \|(\mathbf{I} - \mathbf{P}'_0)\mathbf{B}_{red}\|_F^2 + \|\mathbf{P}'_0 \mathbf{B}_{red}(\mathbf{I} - \mathbf{P}'_0)\|_F^2 \\ &\leq \|\mathbf{U}_{\downarrow m_2+1-r}\|_F^2 + \|\mathbf{V}^H_{\downarrow m_2+1-r}\|_F^2, \end{aligned} \tag{B.7}$$

where the $\mathbf{U}_{\downarrow m_2+1-r}, \mathbf{V}_{\downarrow m_2+1-r} \in \mathbb{C}^{(m_2+1-r)\times r}$ contain the last $m_2+1-r$ rows of the SVD factors $\mathbf{U}$ and $\mathbf{V}$ of $\mathcal{S}(\mathbf{p})$ respectively.

Finally, by Lemma Appendix B.1, we have that the right hand side in (B.7) is bounded by

$$\|\mathbf{U}_{\downarrow m_2+1-r}\|_F^2 + \|\mathbf{V}^H_{\downarrow m_2+1-r}\|_F^2 \leqslant 2\|\mathbf{U}'_{\downarrow m_2+1-r}\|_F^2 = \|\mathbf{P}' - \mathbf{P}_0\|_F^2$$

where $\mathbf{U}' \in \mathbb{C}^{(m_2+1) \times r}$ is the left factor of the SVD of $\mathcal{S}'(\mathbf{p})$ defined in (B.5), and the last equation is from [20, eqn. (29)] and the fact $\mathbf{P}' = \mathbf{U}'(\mathbf{U}')^H$. Hence, the proof is complete. □

*Proof of Corollary 4.3.* The case $|\rho| = 1$ is easy. Consider the case $|\rho| \neq 1$. It is evident that $C(\rho) < 1$ if and only if

$$|\rho^{m+1} - \rho^{-(m+1)}| < y.$$

Denote $x = |\rho|^{m+1} > 0$. If $|\rho| > 1$, then the condition becomes

$$x - \frac{1}{x} < y \iff x^2 - yx - 1 < 0 \iff x < \frac{\sqrt{y^2+4}+y}{2}.$$

If $|\rho| < 1$, the condition becomes

$$\frac{1}{x} - x < y \iff x^2 + yx - 1 > 0 \iff x > \frac{\sqrt{y^2+4}-y}{2}.$$

The statement follows from the properties of the logarithm. □

*Proof of Proposition 4.6.* Consider the change of variables (25). Then it is obvious that (26) is equivalent to (27). Indeed, it follows from (24) and from the fact that

$$\|\mathbf{p}_{sc,(1:n)} - \mathbf{p}_{sc,0}\|_{W_3} = \|\mathbf{p}_{(1:n)} - \mathbf{p}_0\|_{W_2}.$$

The scaling (28) is needed to perform the change of variables that is inverse to (25). □